\documentclass[times]{TRR}

\usepackage{moreverb,url}
\usepackage{xr-hyper}
\usepackage[colorlinks,linkcolor=black,bookmarksopen,bookmarksnumbered,citecolor=black,urlcolor=black]{hyperref}
\usepackage{minitoc}
\usepackage[T1]{fontenc}
\usepackage[utf8]{inputenc}

\newcommand\BibTeX{{\rmfamily B\kern-.05em \textsc{i\kern-.025em b}\kern-.08em
T\kern-.1667em\lower.7ex\hbox{E}\kern-.125emX}}
\newcommand\independent{\protect\mathpalette{\protect\independenT}{\perp}}
\def\independenT#1#2{\mathrel{\rlap{$#1#2$}\mkern2mu{#1#2}}}

\renewcommand{\textbf}[1]{\begingroup\bfseries\mathversion{bold}#1\endgroup}

\newtheorem{assumption}{Assumption}

\makeatletter
\newcommand*{\addFileDependency}[1]{
 \typeout{(#1)}
 \@addtofilelist{#1}
 \IfFileExists{#1}{}{\typeout{No file #1.}}
}

\makeatletter
\DeclareRobustCommand*{\nameref}{%
\color{black}%
        \@ifstar\T@nameref\T@nameref
        }%
\makeatother

\makeatother
\newcommand*{\myexternaldocument}[1]{
    \externaldocument{#1}
    \addFileDependency{#1.tex}
    \addFileDependency{#1.aux}
}

\myexternaldocument{Supplementary_materials}

\begin{document}

\runninghead{Nevoret, Katsahian and Guilloux}

\title{Debiasing the estimate of treatment effect on the treated with time-varying counfounders
}

\author{Camille Nevoret\affilnum{1,2,3}, Sandrine Katsahian\affilnum{3,4} and Agathe Guilloux\affilnum{1}}

\affiliation{\affilnum{1} Université Paris-Saclay, CNRS, Univ Evry, Laboratoire de Mathématiques et Modélisation d'Evry, 91037, Evry-Courcouronnes, France.\\
\affilnum{2} CRC, Université de Paris, INRIA, EPI HEKA, INSERM UMR 1138, Sorbonne Université, France. \\
\affilnum{3} CEMKA, 43 Bd Maréchal Joffre, 92340, Bourg La Reine, France. \\
\affilnum{4} Inserm, Centre d'Investigation Clinique 1418 (CIC1418) Epidémiologie Clinique, Paris, France.
}

\corrauth{Camille Nevoret}

\begin{abstract}
With the increased availability of large health databases comes the opportunity of evaluating treatment effect on new data sources. Through these databases time-dependent outcomes can be analysed as events that can be measured using counting processes. Estimating average treatment effect on the treated (ATT) requires modelling of time-varying covariate and time-dependent treatment and outcome. Gran et al. proposed an easy-to-implement method based on additive intensity regression to estimate ATT. We introduce a debiased estimate of the ATT based on a generalization of the Gran’s model for a potentially repeated outcome and in the presence of multiple time-dependent covariates and baseline covariates. Simulation analyses show that our corrected estimator outperforms Gran's uncorrected estimator. Our method is applied to intensive care real-life data from MIMIC-III databases to estimate vasoppressors effect on patients with sepsis.
\end{abstract}

\maketitle

\section{Introduction} \label{Intro}

Health data warehouse are increasingly available and could be an alternative to the prospective cohort studies traditionally designed for epidemiological research because data are immediately available and projects are thus less time consuming and expensive to run. In most data sources such as health registries or electronic health records, patients are followed over time, resulting in longitudinal data. Statistical analyses are therefore performed in the presence of time-dependent covariates and time-varying treatment. The development of longitudinal databases makes the study of recurring events easier such as repeated events or events that can be measured using counting processes. For example, they are useful when studying rehospitalizations, relapses in cancer, asthma attacks, or multiple sclerosis relapses. 

Time-dependent confounding occurs when a covariate is both affected by past treatment and associated with future treatment choice and outcome. Many longitudinal studies aim to estimate the causal effect of time-varying exposure on the outcome. The main issue of these analyses is to handle such confounding. Specific methods address this issue by taking into account time-confounding and thus avoid biased estimates~\cite{daniel_methods_2013,karim_comparison_2018,thomas_matching_2020}. The most common method for dealing with time-dependent confunding is the inverse probability weighted marginal structural Cox model~\cite{fewell_controlling_2004}. Instrumental variable methods~\cite{hogan_instrumental_2004} or methods derived from g-methods can also be used.

Two types of treatment effect estimator, the average total treatment effect (ATE) and the average treatment effect on the treated (ATT), can be identified. The most commonly used, the ATE, aims identifying the treatment effect in the whole population, eligible for treatment. It allows to answer the question "What would happen if all the patients had received the evaluated treatment, at every timepoint as compared with they had received its comparator?". The ATT aims at identifying the average treatment effect only in patients who actually received the treatment of interest. With this method, the effect of treatment is not assessed in  patients who never received the treatment. The choice of estimator depends on causal research question of interest and on its availability for the method used. For example, it may be more interesting to evaluate the treatment effect on the population that received that treatment rather than on the whole population. Indeed, in clinical practice, the prescription of a treatment is motivated by a set of indicators, as patients clinical characteristics or the likelihood that the treatment will be effective for that patient. In addition, depending on the data available, only information on treated patients may be available. 

The ATT can be estimated mainly by g-estimation of structural nested models~\cite{wang_g-computation_2017}. However, this could be complex and prevents from its common use in practice~\cite{vansteelandt_structural_2014}. Gran et al. proposed an easy-to-implement method that allows to estimate the treatment effect in treated patients and to study intermediate time-dependent covariate trajectories~\cite{gran_estimating_2016}. This model was presented for end-point outcome and an unique time-varying covariate. It consists of two steps. The first step is to model time-varying covariates on the time interval where patients have not yet been treated. Then, the causal treatment effect is estimated by modelling the untreated evolution of time-varying covariates. Finally, treatment effect is estimated using intensity regression models. The proposed method is developed for additive intensity regression models~\cite{aalen_linear_1989}. When studying causality in censored data, this method allows a more explicit approach to processes~\cite{aalen_can_2016} and explicit derivations that are not available for the Cox model~\cite{tchetgen_tchetgen_instrumental_2015, vansteelandt_adjustment_2014, martinussen_estimation_2011, strohmaier_dynamic_2015}. This model seems to be a good alternative to the Cox model~\cite{aalen_does_2015, laan_unified_2003}. 

We propose in this paper to generalize the Gran's model for a potentially repeated outcome and in the presence of multiple time-dependent covariates and baseline covariates. Moreover, we propose a debiased estimate of the ATT based on the error produced during the modelling of the covariates where patients are untreated. Simulation analyses show that the corrected estimator that we propose outperforms Gran's uncorrected estimator.

The paper is organized as follows. Section~\nameref{section:model_notations} introduces the different notations and assumptions used in the rest of the paper. Section~\nameref{treatment_estimation} presents the additive intensity regression model and the proposed correction for a general time-dependent covariates modelling method. Section~\nameref{Section_VAR} specifies the correction in case of vector autoregressive model. Section~\nameref{simulation} summarises results from simulations. Section~\nameref{Application} presents the results of the models on intensive care real-life data from MIMIC-III data base~\cite{johnson_mimic-iii_2016}.

\section{Model and notations} \label{section:model_notations}
\subsection{Model}

 Consider an  individual who suffers from a specific disease with a follow-up over time. At time $S$, he/she initiates his/her treatment. We assume that once started the treatment is continued, so that the treatment process $D$ is null before $S$ and equals $1$ after $S$. Together with the treatment, we observe $d_Z$-dimensional baseline covariates $Z=\big(Z^{1}, \ldots,Z^{d_Z}\big)$ and time-dependent $d_X$-dimensional covariates $t \mapsto X(t) = (X^1(t), \ldots, X^{d_X}(t))$. We focus on pathologies for which the outcome of interest can be measured via a counting process $N$ ($N(t)$ being the number of events recorded in $[0, t]$), see examples in~\nameref{Intro} and~\nameref{Application} sections. We assume also that $N$ verifies the Aalen additive intensity model~\cite{aalen_linear_1989}. 

\begin{assumption}
\label{assump:additive_mean}

With respect $\big(\mathcal F^\star_{t}\big)_{t \geq 0}$,  the historical filtration spanned by $Z$, $X$, $D$, and $N$, we assume the process $X$ is predictable and that $N$ has the following intensity: 
\begin{equation*} 
    \mu^*(t) = \alpha_0^*(t) + \alpha_Z^*(t) Z + \alpha_X^*(t) X(t) + \alpha_D^*(t) D(t).
\end{equation*}
\end{assumption}

As in~\cite{gran_estimating_2016}, we will assume further for simplicity in what follows that independent censoring can occur. Censoring can be assumed to be independent when analyzing data from intensive care unit or administrative databases but not in all pathologies database. Methods allowing dependent censoring such as adjustment using inverse probability of censoring weighting should be used~\cite{jewell_recovery_1992}.

\subsection{Counterfactual quantities, assumptions}
We now consider that we can intervene on the treatment decision at time $S$. Given the definitions introduced above, we have  with probability $1$
\begin{align*}
    D(S) = 1, \;
    D(s) = 0 \text{ for } s < S \text{ and }\;
    D(t) = 1 \text{ for } t \geq S.
\end{align*} However assuming that the positivity assumption, see e.g.~\cite{hernan2009observation} for example, is fulfilled implies that 
\begin{equation*}
    0 < \mathbb P\big(D(t) = 1 | D(t-)=0, \mathcal F_{t-}\big) < 1.
\end{equation*} At any time $t$ before treatment initiation, it can be started or not with a positive probability. This excludes certain pathologies. In practice the counterfactual reasoning requires this positiving assumption which has to be checked on the data.

In the hypothetical situation where the treatment is not initiated at $S$, we denote by $X^{0|S}$ the covariate process after $S$. This counterfactual process corresponding to what would have been observed after $S$. It is important to remark that it cannot be observed, so that its values will have to be estimated (see Section~\nameref{Section_VAR}). The potential covariate process in the situation where the treatment is initiated at $S$ is denoted by $X^{1|S}$.

We also introduce the two potential counting processes $N^{1|S}$ recording the event history after $S$ when the treatment is actually initiated at $S$ and $N^{0|S}$ when the treatment is not initiated. Their respective intensities are denoted by $\mu^{1|S}$ and $\mu^{0|S}$.

The consistency assumption in this case, see~\cite{hernan2009observation}, implies that for all $t \geq S=s$
\begin{equation*}
    X^{1|S=s}(t) = X(t) \text{ and } N(t) = N^{1|S=s}(t).
\end{equation*} We also assume conditional exchangeability conditionally to $S=s$, for all $a=0,1$, all $t\geq s$ and conditionally to the past $\mathcal F_{s-}$:
\begin{align*}
    &\mathbb E(dN^{a|S=s}(t) |  D(s), \mathcal F_{s-}) = \mathbb E(dN^{a|S=s}(t) |  \mathcal F_{s-}) \text{ and }\\
    &\mathbb E(X^{a|S=s}(t) |  D(s), \mathcal F_{s-}) = \mathbb E(X^{a|S=s}(t) |  \mathcal F_{s-}).
\end{align*}
This means that at time $S=s$ conditionally to the past $\mathcal F_{s-}$, if two perfectly similar people (with exactly the same history of treatment and exposure) one receiving the treatment, the other one not receiving the treatment, have a counterfactual outcome which is the other one actual observed outcome. Under Assumption~\eqref{assump:additive_mean}, this can be rewritten, for $t\geq S=s$, as
\begin{align*}
     \mu^{1|S=s}(t) &= \alpha_0^*(t) + \alpha_D^*(t) + \alpha_Z^* Z + \alpha_X^*(t) X^{1|S=s}(t) \text{ and }\\
     \mu^{0|S=s}(t) &= \alpha_0^*(t) + \alpha_Z^* Z + \alpha_X^*(t) X^{0|S=s}(t).
\end{align*}

 The exchangeability assumption can hold only if no unmeasured covariate influences our model. This is a strong assumption in practice but is common in causality reasoning, see e.g.~\cite{gran_estimating_2016, hernan2009observation} among many others. Following~\cite{hernan2009observation}, the clinician opinion expertise is key in proposing a list of covariates associated with treatment indication to be included in the baseline covariates $Z$ and the time dependant covariates process $X$ so that this assumption is fulfilled at least approximately conditional on all variables recorded.

A stronger assumption can be found in the literature~\cite{hernan2009observation, laan_unified_2003}: conditionally to $S=s$, for all $a=0,1$ and conditionally to the past $\mathcal F_{s-}$ 
\begin{equation*}
    X^{a|S}(t) \independent D(s) \text{ and } N^{a|S}(t) \independent D(s), \text{ for all } t\geq s.
\end{equation*}

\subsection{ATT definition and causal estimate}

We assume an additive intensity regression model for the outcome processes. Under the model and the assumptions presented in previous paragraph, the time varying causal intensity difference can be then defined as
\begin{align*}
  d^*(t,s) &=  \mu^{1|S=s}(t) - \mu^{0|S=s}(t)\\
  &= \alpha_D^*(t) D(t)+ \alpha_X^*(t)  (X^{1|S=s}(t) -X^{0|S=s}(t) ).
\end{align*}
With these definitions, we can rewrite the intensity of $N$ as
\begin{equation*}
     \mu^*(t) = \alpha^*_0(t) + \alpha^*_Z Z + \alpha^*_X(t) X^{0|S=s}(t) + d^*_i(t, s)D(t),
\end{equation*} 
and, for $t \geq S$, the time varying ATT as
\begin{multline*}
    \text{ATT}(t) = \mathbb E[ d^*(t,S)|t \geq S] = \alpha_D^*(t) + \\
    \alpha_X^*(t) \mathbb E\big[X^{1|S}(t) -X^{0|S}(t)\big| t \geq S].
\end{multline*}

This has been described in~\cite{gran_estimating_2016} and strongly relies on the Aalen additive intensity model of Assumption~\eqref{assump:additive_mean}. We refer the reader to~\cite{aalen_does_2015} for more discussions on that matter.

\section{Additive intensity model for the treatment effect estimation}
\label{treatment_estimation}

Our data consists in the observation for $n$ independent individuals $i = 1, \ldots n$ of $Z_i$, $X_i$, $S_i$, $D_i$ and $N_i$. Each individual is followed between $t = 0$ and $t = \tau_i$ which can be the end time of the study. The intensity of $N_i$ can be written as: 
\begin{align*}
    \mu^*_i(t) &= \alpha^*_0(t) + \alpha^*_Z Z_i + \alpha^*_X(t) X_i^{0|S=S_i}(t) + d^*_i(t,S_i)D_i(t) \\
    &= W_i(t)^{\top}A^*(t).
\end{align*}

We can recognize the writing of standard regression where treatment effect is adjusting on treatment, the baseline covariates and the counterfactual covariates, $X^{0|S=S_i}_i(t)$. We remark that in that case the ATT is the treatment regression coefficient.

\subsection{Identification of the treatment effect estimate}

The aim of these section is to define an estimator of $A^*$ in the hypothetical situation, false in practice, where the trajectories of the counterfactual covariate processes $X_i^{0|S=S_i}$ for $i=1, \ldots,n$ and $t \geq S_i$, or equivalently the processes $W_i$, would be observed. We simply show that the traditional empirical least-squares risk (see e.g.~\cite{gaiffas_high-dimensional_2012}) for the additive intensity model would lead to a correct estimator of $A^*$. Let 
\begin{equation*}
    A(t) = (\alpha_0(t),\alpha_Z,\alpha_X(t),d_i(t,S_i))^{\top} \text{, } \forall t \geq 0,
\end{equation*} be a candidate for the estimation of $A^\star$. The squared risk of $A \in \mathbb{R}^{d_X+d_Z+2}$ is defined as, see~\cite{gaiffas_high-dimensional_2012}:
\begin{align}\label{eqn:classical_risk}
  r_n(A) &= \frac{1}{n}\sum^n_{i = 1}\int_{0}^{\tau_i} A(t)^{\top} W_i(t) W_i(t)^{\top} A(t)dt
    \\&- \frac{2}{n}\sum^n_{i = 1} \int_{0}^{\tau_i} A(t)^{\top} W_i(t)dN_i(t).\nonumber
\end{align}
The fact that 
\begin{equation}
\label{eqn:A_hat}
   \widehat{A} = \text{arg}\min\limits_{A} r_n(A) 
\end{equation} is a good candidate for the estimation of $A^*$ can be checked by computing the expectation of $r_n(A)$. Calculations presented in Supplementary Materials, section 1.1, allow us to deduce that $A^*$ can be expressed as:
\begin{equation*}
    A^* = \text{arg}\min\limits_{A} \mathbb{E}[r_n(A)],
\end{equation*} which  shows that $\widehat{A}$ is an estimator of $A^*$.

\subsection{Estimation of counterfactual covariates}

In our situation, the counterfactual trajectories $X_i^{0|S = S_i}$ cannot be observed, as in the data, patient $i$ is treated as time $S_i$. To estimate $A^\star$, we need to estimated these trajectories. We give an exemple of such estimations in Section~\nameref{Section_VAR}. We consider for now that we can model the trajectories and estimate the values of $X^{0|S=S_i}(t)$ for $t \geq S_i$ by $\widetilde{X}^{0|S=S_i}_i(t)$. Being an estimation, $\widetilde{X}^{0|S=S_i}_i(t)$ is not equal to $X^{0|S=S_i}(t)$ but we assume that we can write:
\begin{align}
\label{eqn:Wtilde_W}
    \widetilde{X}^{0|S=S_i}_i(t) &= X^{0|S=S_i}_i(t) + \epsilon_i(t), \text{ for all } t \geq S_i,
\end{align}
where $\epsilon_i(t)$ is centered and has finite variance. We finally denote by $\widetilde{W}_i(t)$ the vector $W_i(t) + \xi_i(t)$ with $\xi_i(t) = (0,0_{\mathbb{R}^{d_z}},\epsilon_i(t),0)$ for $t \geq S_i$ and $\xi_i(t) = (0, \ldots, 0)$ for $t < S_i$.

Mimicking the computation in the above paragraph, let us introduce $\widetilde{R_n}(A)$ the squared risk of $A$ using counterfactual covariates estimate $\widetilde{W}_i(t)$:
\begin{align*}
    \widetilde{R_n}(A) = \frac{1}{n}\sum^n_{i = 1}\int_{0}^{\tau_i} A(t)^{\top} \widetilde{W}_i(t) \widetilde{W}_i(t)^{\top} A(t)dt \\
    - \frac{2}{n}\sum^n_{i = 1} \int_{0}^{\tau_i} A(t)^{\top} \widetilde{W}_i(t)dN_i(t).
\end{align*} Computations detailed in Equation (1) of Supplementary materials allow to write  $\widetilde{R_n}(A)$ in function of  $r_n(A)$ using Equation~\eqref{eqn:Wtilde_W} as 
\begin{align} \label{eqn:Rn_W}
    r_n(A) &= \widetilde{R}_n(A) \\&+  \frac{1}{n} \sum^n_{i = 1} \int_{0}^{\tau_i} \alpha_X(t)^{\top} \epsilon_i(t) \epsilon_i(t)^{\top} \alpha_X(t)dt. \nonumber
\end{align} Using Equation~\eqref{eqn:Rn_W}, the expectation of $\widetilde{R_n}$ is given by the following expression:
\begin{multline}
\label{eqn:identify_error}
    \mathbb{E}[\widetilde{R_n}(A)] =  \mathbb{E}[r_n(A)]  \\ - \frac{1}{n} \sum^n_{i = 1} \int_{0}^{\tau_i} \alpha_X(t)^{\top} \mathbb{E}[\epsilon_i(t) \epsilon_i(t)^{\top}] \alpha_X(t)dt.
\end{multline} Hence $\text{arg}\min\limits_{A} \mathbb{E}[\widetilde{R}_n(A)] \ne A^*$. 

In that sens, the proposal of Gran et al.~\cite{gran_estimating_2016} is biased. We will introduce in the following section a novel estimator of $A^*$ (and the ATT) taking into account the counterfactuals modelling error .

\subsection{Debiased treatment effect estimate}

We showed that minimizing $\widehat{R_n}$ would had to a biased estimated. We propose in this paragraph to correct this bias. Equation~\eqref{eqn:identify_error} implies that 
\begin{equation*}
    \widehat{R_n}(A) =  \widetilde{R_n}(A) + \frac{1}{n} \sum^n_{i = 1} \int_{0}^{\tau_i} \alpha_X(t)^{\top} \widehat{\mathbb{E}[
\epsilon_i(t) \epsilon_i(t)^{\top}]} \alpha_X(t)dt
\end{equation*} is the correct loss to estimated $A^\star$. However, as the unknown expectations $\mathbb{E}[\epsilon_i(t)\epsilon_i(t)^{\top}]$ appear, it cannot be directly used. For some models for the $X_i^{0|S = S_i}$ trajectories, these expectations can however be estimated. This is the case with the vector autoregressive model presented in the next section. Now considering that we have access to such an estimation $\widehat{\mathbb{E}[\epsilon_i(t) \epsilon_i(t)^{\top}]}$ for all $i$ and $t \geq S_i$, we can defined our loss as: 
\begin{equation*}
\widehat{R_n}(A) =  \widetilde{R_n}(A) + \widehat{\text{bias}}(A),
\end{equation*} with \begin{equation}
\begin{split}
    \widehat{\text{bias}}(A) = \frac{1}{n} \sum^n_{i = 1} \int_{0}^{\tau_i} \alpha_X(t)^{\top} \widehat{\mathbb{E}[
\epsilon_i(t) \epsilon_i(t)^{\top}]} \alpha_X(t)dt.
\end{split}
\end{equation}

Finally, the debiased estimator of $A^*$ is given by:
\begin{align}
\label{eqn:A_hat_debiased}
\widehat{A}_{\text{debiased}} &= \text{arg}\min\limits_{A} \widehat{R_n}(A) \nonumber \\
    &= \text{arg}\min\limits_{A} \frac{1}{n} \sum^n_{i = 1} \int_{0}^{\tau_i} A(t) 
\left[
\widetilde{W}_i(t) \widetilde{W}_i(t)^{\top} \right.  \nonumber \\
&\quad -  \left. \begin{pmatrix}
0 & \cdots & 0 \\
\vdots & \widehat{\mathbb{E}[\epsilon_i(t) \epsilon_i(t)^{\top}] } & \vdots  \\
0 &  \cdots & 0
\end{pmatrix} \right] A(t)
dt \nonumber \\
&-\frac{2}{n} \sum^n_{i = 1} \int_{0}^{\tau_i} A(t)\widetilde{W}_i(t)dN_i(t). 
\end{align}

We now propose a simple model for the continuous time-varying $d_x$-dimensional covariate. $X_i$ in which the estimation $\widehat{\text{bias}}(A)$ is easily computable.

\section{Example with a vector autoregressive model VAR(1)}
\label{Section_VAR}

\subsection{Explicit writing of the bias}

In the following, the observation period from $0$ to $\max\limits_{i} \tau_i$ is split into observed time intervals defined by following times $t_0 = 0, \ldots, t_k, \ldots t_K = \max\limits_{i} \tau_i$. We assume that the time varying covariates are constant over these intervals such that for all $i$, $k$ and $t \in [t_k, t_{k+1})$, $X_i(t) = X_i(t_k)$, and by extension $W_i(t) = W_i(t_k)$. 

Considering the case of continuous time-varying covariates, the simplest model is the vector auto-regressive model~\cite{lutkepohl_introduction_1993}. We explicit the computation in the case of the VAR(1) model:
\begin{equation*}
    X_i(t_k) = \beta_0^{i \star} + \Pi^{\star} X_i(t_{k-1}) + \omega_i(t_k),
\end{equation*}
for all $i$ and $t$, where $\Pi^{\star}$ is the coefficient matrix and $\omega_i$ is an unobservable zero mean white noise vector process with time invariant covariance matrix $\Sigma^{\star}$.

Following the literature on VAR model, the prediction of $X(t_{k+l})$ using the past up $t_k$ is given iteratively by:
 \begin{align*}
     \widetilde{X}_i(t_{k+l}) &= \beta_0^{i\star} + \Pi^{\star} \widetilde{X}_i(t_{k+l-1}) \\
                        &= \sum_{j=0}^{l-1} {\Pi^{\star}}^{j}\beta_0^{i \star} + {\Pi^{\star}}^{l} X_i(t_k).
 \end{align*}
The modelling error can be expressed as follow:
 \begin{equation} 
 X_i(t_{k+l}) - \widetilde{X}_i(t_{k+l}) = \sum_{j=0}^{l-1}  {\Pi^{\star}}^j \omega_i(t_{k+l-j}),
 \end{equation}
with $\Pi^0 = \mathbf{I}_{d_X}$ the identity matrix of size $d_X$. The mean squared error (MSE) matrix for $ \widetilde{X}_i(t_{k+l}) $ is written as:
\begin{align}
    \Sigma(l) &= \text{MSE}(X_i(t_{k+l}) - \widetilde{X}_i(t_{k+l})) 
            = \sum_{j=0}^{l-1} {\Pi^{\star}}^j \Sigma {\Pi^{\star}}^{\top j}. \label{eqn:MSE}
\end{align}

The parameters of the VAR(1) process are estimated using multivariate least squares~\cite{lutkepohl_new_2005}. Estimated parameters is noted $\widehat{\Pi}$ and $\widehat{\beta_0^i}$. The best linear predictor of $X_i(t_{k+l})$ becomes:\begin{equation}
\label{eqn:counterfactual_model}
    \widehat{\widetilde{X}}_i(t_{k+l}) = \widehat{\beta_0^i} + \widehat{\Pi} \widehat{\widetilde{X}}_i(t_{k+l-1}) .
\end{equation}Finally, the estimated MSE matrix of the 1-step forecast is computing using equation \eqref{eqn:MSE} as:
\begin{equation*}
    \widehat{\Sigma}(l) = \sum_{j=0}^{l-1} \widehat{\Pi}^j \widehat{\Sigma} \widehat{\Pi}^{\top^j}.
\end{equation*}

\subsection{Algorithm}
\label{Algorithm}

Our algorithm combines two step.

\subsubsection{First step: counterfactual estimates }
    \begin{enumerate}
        \item  After filtering, only observations where patients are\textbf{ not treated} are kept.
        \item For each time-varying covariates $j$, the matrix in which $t_{\max}^i$ is the last time without treatment for the patient $i$ is designed, i.0.e the discret observe time before $S_i$.
        \item The linear regression with the design matrix of Equation \eqref{eqn:desig_matrix} and the vector of predicted outputs $(X_1^j(1), \cdots, X_1^j(t_{\max}^1), \cdots, X_n^j(1), \cdots, X_{n}^{j}(t_{\max}^n))$ is computed.
        \item Thus we get the estimates of $(\Sigma_{jj})$, $(\Pi_{j,q})_{1 \leq q \leq d_X+d_Z}$ and $\beta_0^i$.
        \item  Counterfactuals are modelled following Equation~\eqref{eqn:counterfactual_model} using estimates of $\Pi$ and $\beta_0^i$, and are denoted $\widehat{\widetilde{X_i}}(t_k)$ for $i=1,\ldots,n$ and $t_k\geq t_i$.
 \end{enumerate}

   \begin{equation}\label{eqn:desig_matrix}
    \begin{pmatrix}
            1 & Z_1^1 & \cdots & Z_1^{d_Z} & {\scriptstyle X_1^1(0)} & \cdots & {\scriptstyle X_1^{d_X}(0)} \\
            \vdots & \vdots &  \vdots &  \vdots & \vdots &  \vdots &  \vdots \\
            1 & Z_1^1 & \cdots & Z_1^{d_Z} & {\scriptstyle X_1^1(t_{\max}^1-1)} & \cdots & {\scriptstyle X_1^{d_X}(t_{\max}^1-1)} \\
            \vdots &\vdots &  \vdots &  \vdots & \vdots &  \vdots &  \vdots \\
           1 &  Z_n^1 & \cdots & Z_n^{d_Z} & {\scriptstyle X_n^1(0)} & \cdots & {\scriptstyle X_n^{d_X}(0)} \\
           \vdots & \vdots &  \vdots &  \vdots & \vdots &  \vdots &  \vdots \\
            1 & Z_n^1 & \cdots & Z_n^{d_Z} & {\scriptstyle X_n^1(t_{\max}^n-1)} & \cdots & {\scriptstyle X_n^{d_X}(t_{\max}^n-1)} \\
            \end{pmatrix}
         \end{equation}
        
 \subsubsection{Second step: ATT estimate with modelled counterfactuals}
    
    The ATT estimate is implemented in different analysis softwares such as the \texttt{aalen} function of the \texttt{timereg} package~\cite{martinussen_dynamic_2006,scheike_analyzing_2011}. We implemented our own method to apply our correction. 
    
    \begin{itemize}
      \item The ATT is estimating by solving Equation~\eqref{eqn:A_hat} and~\eqref{eqn:A_hat_debiased}. 
    \end{itemize}

\section{Simulation study}
\label{simulation}

\begin{figure}[h!]
\centering
\includegraphics[scale=0.7]{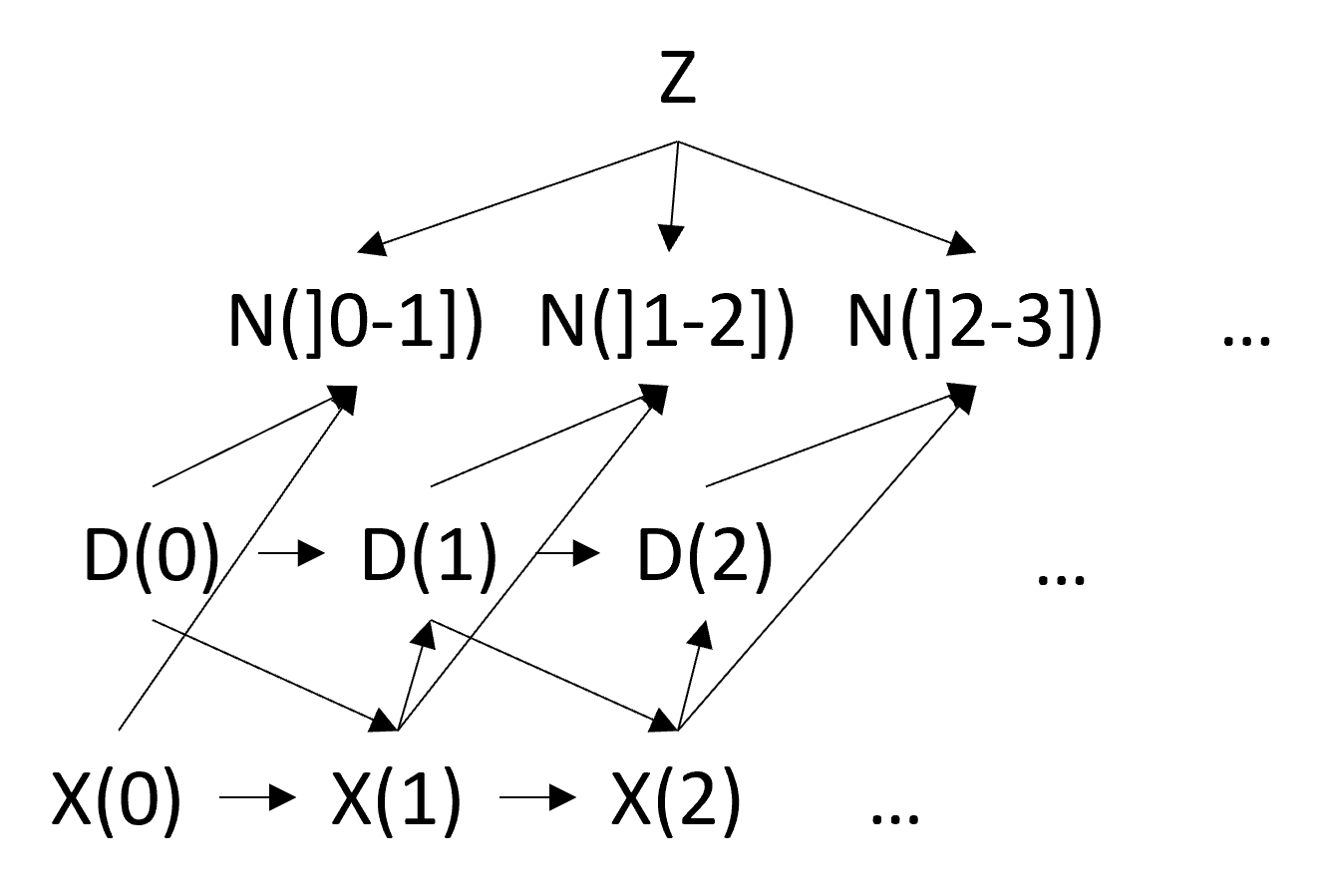}
\captionof{figure}{Directed Acyclic Graph} 
\label{fig:DAG}
\end{figure}

In this section, ATT estimations with and without correction are compared using simulations in the VAR(1) model. This simulation study is based on the simulation setup proposed in~\cite{gran_estimating_2016} . Data are simulated to mimic a cohort analysis where individuals are under risk of having an event of interest which can be prevented or delayed by treatment. Several covariates are simulated over time. Their values depend on treatment status, and reciprocally, treatment initiation depends on these values, see Figure~\ref{fig:DAG}. The probability of having the event is then affected by both the treatment status and the covariate values.

\subsection{Data generation}

Simulation setup is motivated in~\cite{gran_estimating_2016} by an application on the effect of antiretroviral therapy (HAART) on the time to AIDS onset or death. We have extended these simulations with several time-varying covariates and baseline covariates. 

The data are simulated on time intervals $[t_k, t_{k+1})$ with pre-specified time points $t_k = 0,1, .... 11$. On each time interval, we define event status $N(t_k)$, treatment status $D(t_k)$, $d_X$ time-varying covariates, $X(t_k) = (X^1(t_k), \dots, X^d(t_k))$ and $d_Z$ baseline covariates, $Z$. Dataset with one, three and six time-varying covariates are simulated.

At initiation, all patient are untreated. Initial time-varying covariates are simulated as uniform distribution. Three baseline covariates are simulated to mimic age as numbers uniformly distributed, gender as a binomial distribution and a comorbidity score as a Poisson distribution. 

On the interval where the treatment has not yet been initiated, each covariates $X^j(t_k)$ will decrease steadily according to the following equation $X(t_{k+1}) = \kappa_{D0}X(t_k) + \mathcal{N}(0_{\mathbb{R}^d},\Sigma)$ with $\kappa_{D0}$ a $d_X \times d_X$ matrix and $\Sigma = \{ \sigma_{jj} \}_{1 \leq j \leq k}$. While from treatment initiations, $X^j(t_k)$ increase over time following the equation $X(t_{k+1}) = X_i(t+1) = X_i(t) + \kappa_{D1}(\sqrt{1000} - X_i(t) ) + \mathcal{N}(0_{\mathbb{R}^d},\Sigma)$ with $\kappa_0$ a $d_x$-dimensional vector and $\kappa_{D1}$ a $d_X \times d_X$ matrix. Treatment initiation $D(t_k)$ is simulated according to a Bernoulli distribution with parameter $m \sum_{j=1}^{d}\lambda_j e^{\sum_{j=1}^{d} \lambda_j X^j(t_k)})$. We assume that once treatment is initiated, the patient remains on treatment until the end of the follow-up.

Finally, treatment status and time-varying covariates act on the number of experienced events, $N(]t_k, t_{k+1}])$ experienced on $]t_k, t_{k+1}]$ . The event times are simulated according to a non-homogeneous Poisson process with intensity function $\delta D(t_k) + \delta_0 +\sum_{j=1}^{d_Z} \delta_{Zj} Z^j + \sum_{j=1}^{d}\delta_{Xj} X^j(t_k)$. This process is generated using acceptance-rejection method called thinning~\cite{lewis_simulation_1979}. All patients are followed up until time $t_k=11$. The relation between covariates, treatment and outcome is summarized by the direct acyclic graph of Figure~\ref{fig:DAG}.

At the same time, the "true" counterfactual values are simulated by following the evolution of the covariates when $D(t_k) = 0$. The simulation equation is as follows: $\widetilde{X}(t_{k+1}) = \kappa_{D0}\widetilde{X}(t_k) + \mathcal{N}(0_{\mathbb{R}^d},\Sigma)$ whatever the treatment status. The simulation algorithm is presented in Supplementary Materials, see Section 2.

The number of time-varying covariates varied in the simulated framework. The following results are based on simulations with 1 time-varying covariate affecting $D$ and $N$, and simulations with three and six covariates with $d/3$ covariates affecting $D$ and $N$, $d/3$ covariates affecting only $D$ and $d/3$ covariates affecting only $N$. Finally, the parameter $\Sigma_{jj}$ was set to the values $0.4$, $0.8$, $1.2$ and $1.6$. The values of parameters used in the simulations are given in the appendix, see Section~\ref{simulation_algorithme}. We ran the simulations 100 times with the same parameters on data sets of 1000 individuals. All simulations were done in R version 4.1. Codes used for the simulations are available on gitlab~\url{https://gitlab.com/camille.nevoret/att-estimation}.

\subsection{Analysis}

As presented in the section \nameref{Algorithm}, the first step in the simulation study is the modelling of counterfactuals. Within the VAR(1) model, we estimated $\Sigma$ and $\kappa_{D0}$, used to calculate the correction of the ATT estimator.

The data set of the 100 simulation repetitions with the true counterfactual values is used to evaluate the "true" ATT value by Monte Carlo. At each time $t$, we note $d^{MC}(t)$ the "true" ATT. All ATT estimates for each of the 100 simulations are compared to this value. We use the mean integrated square error (MISE) to calculated the error between ATT estimate and the true ATT. For a given estimator $\hat{d^l}(t)$ for the l-th simulation, the MISE is given by: 
\begin{equation*}
    \text{MISE}^l = \mathbb{E}\left[\int^{\tau = 11}_0 (\widehat{d}^l(t) - d^ {MC}(t))^2 dt\right].
\end{equation*}

For each of the 100 simulations, we estimate ATT using five different estimators. 
\begin{itemize}
    \item The first estimator is calculated using the \texttt{aalen} function of the R \texttt{timereg} package. Counterfactual data use for this estimate are the one simulated as "real counterfactuals". Thus, there are no errors related to the modelling of counterfactuals introduced in the ATT estimation. In this sense, the other estimators, based on modelled counterfactuals cannot make more accurate estimations than this one. It is thus uses as a reference. 
    \item Then, two uncorrected estimators are presented. The first one corresponds to the one presented in~\cite{gran_estimating_2016} using \texttt{aalen} function and modelled counterfactuals. The second one is obtain using algorithm presented section \nameref{Algorithm}. 
    \item Finally, two corrected estimators are presented. They use both the algorithm presented section \nameref{Algorithm} and modelled counterfactuals. The first one use the true time invariant covariance matrix, $\Sigma$, and the true $\kappa_{D0}$. In real life data, they are estimated when modelling the counterfactuals. This is done in the second estimator.
\end{itemize}

Wilcoxon tests for the difference in distribution are performed between the corrected estimator with estimated parameters and each uncorrected estimators. 

\subsection{Simulation results}

Table \ref{tb:result_gran_vs_new} shows the mean integrated square error for three ATT estimates: the two uncorrected estimators and the corrected estimator using estimate parameters $\widehat{\Sigma}$ and $\widehat{\kappa}_{D0}$. We observe that, whatever the number of covariates, the true mean integrated square error increases with $\Sigma$. Data in bold are the lowest MISEs. The symbol $^{\star}$ represents a significative difference given by the Wilcoxon test. 

For all simulation parameters, ATT estimates using the \texttt{aalen} function~\cite{martinussen_dynamic_2006,scheike_analyzing_2011} and using our recoded algorithm are very close. Our corrected estimator is always better than uncorrected estimated. We notice significant differences between these estimators, for all simulation scenarios with one covariate.

\begin{table}
\footnotesize\sf\centering
    \caption{Mean MISE $\pm$ standard deviation for corrected and uncorrected estimators ($^{\star}$: Significant Wilcoxon test between corrected estimator with estimated parameters and other estimators, boldface character: the smaller MISE of the corrected and uncorrected estimators)}
\label{tb:result_gran_vs_new}
    \begin{tabular}{|l|c|c|c|}
    \hline
        & Corrected  & \multicolumn{2}{|c|}{Uncorrected estimator} \\
    \cline{3-4} 
    &  estimator & With \texttt{aalen} & Recoded  \\
    &  with $\widehat{\Sigma}$ & function & estimate  \\
    
    \hline
    \multicolumn{4}{|l|}{1 Covariate} \\
    \hline
    $\Sigma_{k,k} = 0.4$  &  \textbf{$0.034 \pm 0.038$}    &  $0.045^{\star} \pm 0.047$ & $0.044^{\star} \pm 0.047$  \\
    $\Sigma_{k,k} = 0.8$  & \textbf{$0.136 \pm 0.131$}   &  $0.178^{\star} \pm 0.155$ & $0.177^{\star} \pm 0.155$    \\
    $\Sigma_{k,k} = 1.2$ & \textbf{$0.412 \pm 0.384$} & $0.533^{\star} \pm 0.445$ & $0.527^{\star} \pm 0.443$   \\
    $\Sigma_{k,k} = 1.6$ & \textbf{$0.787 \pm 0.730$} &  $1.026^{\star} \pm 0.853$ & $1.014^{\star} \pm 0.848$   \\
    \hline
    \multicolumn{4}{|l|}{3 Covariates} \\
    \hline
    $\Sigma_{k,k} = 0.4$ & \textbf{$0.040 \pm 0.046$} &  $0.045 \pm 0.051$ & $0.045 \pm 0.051$  \\
    $\Sigma_{k,k} = 0.8$  & \textbf{$0.224 \pm 0.237$} & $0.257 \pm 0.265$ & $0.256 \pm 0.264$    \\
    $\Sigma_{k,k} = 1.2$ & \textbf{$0.518 \pm 0.548$} & $0.608 \pm 0.604$ & $0.604 \pm 0.603$   \\
    $\Sigma_{k,k} = 1.6$  & \textbf{$1.381 \pm 1.299$} & $1.598 \pm 1.434$ & $1.587 \pm 1.429$   \\
    \hline
    \multicolumn{4}{|l|}{6 Covariates} \\
    \hline
    $\Sigma_{k,k} = 0.4$ & \textbf{$0.075 \pm 0.080$} &  $0.081 \pm 0.095$ & $0.081 \pm 0.096$   \\
    $\Sigma_{k,k} = 0.8$ & \textbf{$0.320 \pm 0.360$} & $0.338 \pm 0.376$ & $0.337 \pm 0.377$   \\
    $\Sigma_{k,k} = 1.2$ & \textbf{$0.901 \pm 1.012$} & $0.949 \pm 1.043$ & $0.947 \pm 1.045$   \\
    $\Sigma_{k,k} = 1.6$ & \textbf{$2.448 \pm 2.316$} & $2.567 \pm 2.404$ & $2.560 \pm 2.406$    \\
    \hline
    \end{tabular}
\end{table}

Table \ref{tb:result_new_true} allows to compare our corrected ATT estimate with estimated parameters, which is the one that can be used on real data, to corrected the ATT estimate with known parameters and the estimate using the "real counterfactuals". It appears that MISE of corrected estimates are very close which shows that counterfactual modelling lead to good estimates of $\Sigma$ and $\kappa_{D0}$. We can also note large standard deviations of the order of the means of the MISE. These large standard deviations do not seem to be inherent to our estimation method. Indeed, we observe the same type of deviation for the estimate using "real counterfactuals". 

Finally, for all simulation parameters tested, the corrected estimators appear to behave like the estimator based on the "real counterfactuals" and perform better than the uncorrected estimators.

\begin{table}
\footnotesize\sf\centering
    \caption{Mean MISE $\pm$ standard deviation for corrected  estimators and estimator obtained using "real counterfactuals"}
\label{tb:result_new_true}
    \centering
    \begin{tabular}{|l|c|c|c|}
    \hline
        &  \multicolumn{2}{|c|}{Corrected estimator} & Estimate \\
        &  \multicolumn{2}{|c|}{estimator} &  using "real \\
    \cline{2-3} 
     & with $\widehat{\Sigma}$ &  with $\Sigma$ &  counterfactuals"\\
    
    \hline
    \multicolumn{4}{|l|}{1 Covariate} \\
    \hline
    $\Sigma_{k,k} = 0.4$ & $0.034 \pm 0.038$ & $0.034 \pm 0.038$ &  $0.034 \pm 0.034$\\
    $\Sigma_{k,k} = 0.8$  &  $0.136 \pm 0.131$  & $0.136 \pm 0.131$ &  $0.060 \pm 0.066$ \\
    $\Sigma_{k,k} = 1.2$  & $0.412 \pm 0.384$ &  $0.412 \pm 0.384$ &  $0.186 \pm 0.168$\\
    $\Sigma_{k,k} = 1.6$ & $0.787 \pm 0.730$ & $0.783 \pm 0.726$ &  $0.315 \pm 0.436$ \\
    \hline
    \multicolumn{4}{|l|}{3 Covariates} \\
    \hline
    $\Sigma_{k,k} = 0.4$  & $0.040 \pm 0.046$ & $0.040 \pm 0.046$ & $0.041 \pm 0.044$\\
    $\Sigma_{k,k} = 0.8$ & $0.224 \pm 0.237$ & $0.224 \pm 0.237$  & $0.170 \pm 0.237$ \\
    $\Sigma_{k,k} = 1.2$ & $0.518 \pm 0.548$ & $0.516 \pm 0.546$  & $0.283 \pm 0.379$ \\
    $\Sigma_{k,k} = 1.6$ & $1.381 \pm 1.299$ & $1.371 \pm 1.291$ & $0.652 \pm 0.956$ \\
    \hline
    \multicolumn{4}{|l|}{6 Covariates} \\
    \hline
    $\Sigma_{k,k} = 0.4$  & $0.075 \pm 0.08$ &  $0.075 \pm 0.079$ & $0.071 \pm 0.060$ \\
    $\Sigma_{k,k} = 0.8$  & $0.320 \pm 0.360$ & $0.320 \pm 0.359$ & $0.330 \pm 0.471$ \\
    $\Sigma_{k,k} = 1.2$  & $0.901 \pm 1.012$ & $0.898 \pm 1.011$  & $0.775 \pm 1.033$ \\
    $\Sigma_{k,k} = 1.6$ & $2.448 \pm 2.316$ & $2.437 \pm 2.306$ & $1.563 \pm 2.146$ \\
    \hline
    \end{tabular}
\end{table}

\section{Application}
\label{Application}

Now, let us apply method to real life data from the open access MIMIC-III database~\cite{johnson_mimic-iii_2016}. This database contains deidentified data of patients admitted to the critical care unit of the Beth Israel Deaconess Medical Center in Boston, Massachusetts between 2001 and 2012. Over this period, 53,423 distinct admissions of adult patients were identified. Data includes demographics data, laboratory and microbiology test results, diagnosis and procedure codes, bedside monitoring data (vital signs, waveforms, trends...), fluids, medications. 

The focus here is on patients with sepsis. These patients are identified according to the algorithm defined on~\cite{martin_epidemiology_2003}, based on specific ICD-9 codes and procedures codes. We try to estimate the effect of vasopressor on death. The following vasopressors, norepinephrine, epinephrine, phenylephrine, vasopressin, dopamine, isuprel, are sought in the fluids administered to patients during their stay. Several prognostic factors are consider, systolic and diastolic blood pressures as time-dependent factors and gender, age, lactate, creatinine and SOFA score~\cite{vincent_sofa_1996} at inclusion in intensive care as baseline factors. Blood pressure data are filled in every hour. Patients are followed from entry into intensive care until discharge or death.  

Time-dependent covariates are smoothed using moving average of 5 values and standardized to get more stable trajectories. Observed blood pressures at inclusion for each patients are considered as baseline covariates. Counterfactual covariate trajectories under the scenario of no treatment are estimated according to a VAR model as presented in section~\nameref{Section_VAR}. Estimate starts at treatment time and stops at death or discharge. Baseline variables mentioned above are used as adjusted covariates.  

\begin{figure}
\centering
\includegraphics[scale = 0.35]{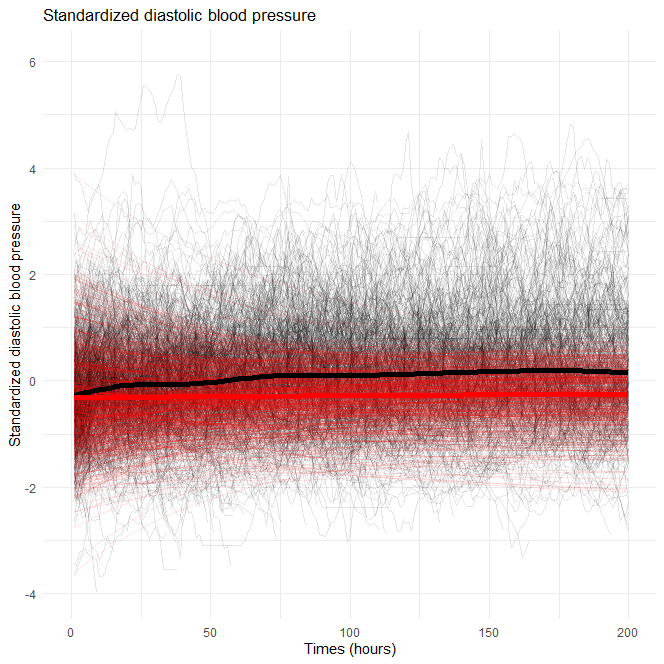}
\caption{Standardized diastolic blood pressure (black: observed trajectoried, red: modelled trajectorie)}
\label{figure_PAD}
\end{figure}

\begin{figure}
\centering
\includegraphics[scale = 0.35]{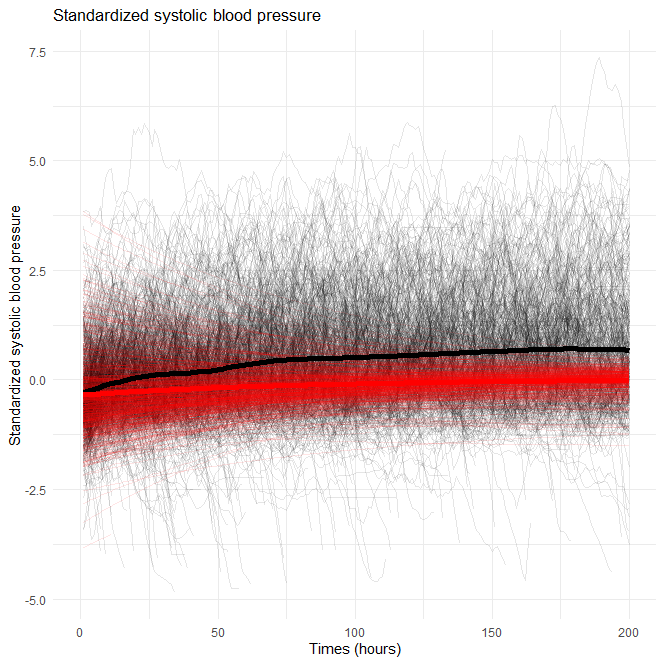}
\caption{Standardized systolic blood pressure (black: observed trajectoried, red: modelled trajectorie)}
\label{figure_PAS}
\end{figure}

Figures~\ref{figure_PAS} and~\ref{figure_PAD} shows observed and counterfactual covariates from treatment initiation. The graph is truncated at 200 hours to keep enough patients. The red lines represent counterfactual covariates under the assumption that the patients would not have been treated. These are unobserved trajectories. The grey lines represent observed covariates when patients are treated. Graphically, we see, for treated individuals, an increase in systolic and diastolic blood pressure which is the clinical effect of vasopressors. 

As a reminder, the objective here is to evaluate the effect of vasopressors on death. ATT is estimated using Gran's estimator and ours. We include in these models the above mentionned covariates: smoothed standardised systolic and diastolic blood pressure grouped into hourly intervals, gender, age, lactate, creatinine, SOFA score, systolic and diastolic blood pressure on entry into intensive care. These estimations are compared to naive estimation and ATE estimation. Naive estimation is obtained using Aalen model with baseline covariates and observed time-dependent covariates. ATE is estimated using marginal structural additive model with the same covariates. Time-dependent covariates where replaced by stabilised inverse probability of treatment weights.

\begin{figure}
\centering
\includegraphics[scale = 0.3]{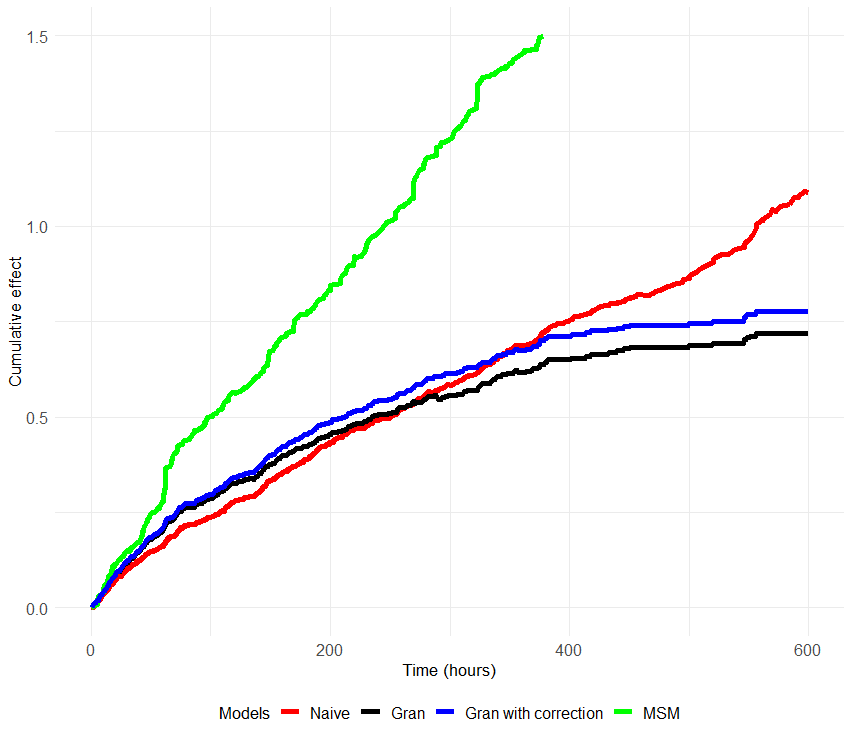}
\caption{Cumulative treatment effect estimated using Naive model, Marginal structural model Gran’s estimator with and without correction.}
\label{figure_ATT}
\end{figure}

Cumulative treatment effect is showed in Figure~\ref{figure_ATT} for vasopressor versus no treatment. Treatment has a positive contribution to the intensity (increasing curve) all the more important in the first days. In other words, treatment have not a protective effect neither on the eligible population (ATE) nor on the patients actually treated (ATT). The direction of the effect can mainly be explained because prescription of vasopressor is directly correlated to the severity of septic shock and thus to death. Moreover these methods do not allow to take into account the dose which may have an effect on death. 

Finally, we observe that the cumulative effect estimate from the naive model and from the MSM seems to be linear, while this effect seems to stabilize for the Gran's estimators and ours. We observe that the correction has an increasing effect with time, which is explained by the increase in the number of patients treated over time.  

\section{Discussion}

In this article, we proposed a corrected estimator of the ATT under time-varying confounders. We were inspired by the method developed by Gran et al.~\cite{gran_estimating_2016}. This method consists of two steps. The first one is the modelling of time-dependent covariates if the patients had not been treated and the second is the estimation of the treatment effect. The correction we propose is based on the error that is made when modelling the counterfactuals. Indeed, as in any modelling, there is a difference between the "real" data and those modelled. 
Our model allows the study of:
\begin{enumerate}
    \item terminal outcomes such as death or repeated events/counting processes such as rehospitalizations or asthma attacks ;
    \item multiple time varying covariates and baseline covariates. 
\end{enumerate}
All simulation scenarios showed that our corrected estimator provides better estimates of the ATT than those obtained with uncorrected estimators. 

Estimates of ATT and ATE answer the two different questions. The ATT estimator addresses the causal question "Is the treatment efficient in patients in whom it is actually prescribed?" rather than "Is the treatment efficient in the whole population" which is addressed by ATE. It may be interesting to estimate these two measures in the same study to get additional information about the effect of the treatment. Their comparison can be used to assess the effect of current treatment policy such as the choice of patients receiving treatment and the treatment initiation time. The methods used to estimate these measures can also be used to select optimal treatment regimes. This is particularly the case for g-estimation method~\cite{barrett_doubly_2014}, Inverse Probability of Censoring Weighting model~\cite{shen_estimation_2017} or a non-parametric structural equation model and g-computation~\cite{diaz_nonparametric_2021}.

Our correction is introduced int the cases where the data are continuous and the counterfactuals are modelled by a VAR(1) model. This model is simple to implement and allows an explicit writing of the correction. Other methods could be used such as models more specific to diseases and time-dependent covariates such as differential equations. However, it should be noted that the more complex the model, the more difficult it will be to make explicit the modelling error and thus the correction of the ATT estimator. 

Note that in this paper, only continuous covariates are studied. Depending on the databases analyzed, the time-dependent covariates may be binary or count data such as the number of nursing or general practitioner visits over each time interval studied. Specific methods to model the counterfactuals for this type of variable should be developed. It would also be interesting to be able to consider different types of data in the same model.

\end{document}


\maketitle

\section{Computations detailed}
\subsection{Expectation of $r_n(A)$}
\label{appendix-11}

\begin{align*}
\mathbb{E}[r_n(A)] &=
     \mathbb{E}[\frac{1}{n}\sum^n_{i = 1}\int_{0}^{\tau_i} A(t)^{\top} W_i(t) W_i(t)^{\top} A(t)dt  - \frac{2}{n}\sum^n_{i = 1} \int_{0}^{\tau_i} A(t)^{\top} W_i(t)dN_i(t)] \\
    &= \frac{1}{n}\sum^n_{i = 1}\int_{0}^{\tau_i} A(t)^{\top} W_i(t) W_i(t)^{\top} A(t)dt - \frac{2}{n}\sum^n_{i = 1} \int_{0}^{\tau_i} A(t)^{\top} W_i(t)\mathbb{E}[dN_i(t)] \\
\intertext{Using additive structure of the intensity of counting process $N_i$ : 
 $\mathbb{E}[dN_i(t)] = \mu_i^*(t)dt = W_i(t)^{\top}A^*(t)dt $, we obtain }
    &= \frac{1}{n}\sum^n_{i = 1}\int_{0}^{\tau_i} A(t)^{\top} W_i(t) W_i(t)^{\top} A(t)dt  - \frac{2}{n}\sum^n_{i = 1} \int_{0}^{\tau_i} A(t)^{\top} W_i(t) W_i(t)^{\top}A^*(t) dt \\
    &= \frac{1}{n}\sum^n_{i = 1}\int_{0}^{\tau_i} (A(t) - A^*(t))  ^{\top} W_i(t) W_i(t)^{\top} (A(t) - A^*(t)) dt \\
    &\quad - \frac{1}{n}\sum^n_{i = 1} \int_{0}^{\tau_i} A^*(t)^{\top} W_i(t) W_i(t)^{\top}A^*(t) dt \\
     &=  ||A-A^*||_n^2 - \frac{1}{n}\sum^n_{i = 1} \int_{0}^{\tau_i} A^*(t)^{\top} W_i(t) W_i(t)^{\top}A^*(t).
\end{align*}

\subsection{$r_n(A)$ in function of $\widetilde{R_n}(A)$}
\label{appendix-12}

\begin{align} \label{eqn:supp_mat_rnA}
r_n(A) &= \frac{1}{n}\sum^n_{i = 1}\int_{0}^{\tau_i} A(t)^{\top} W_i(t) W_i(t)^{\top} A(t)dt \nonumber  - \frac{2}{n}\sum^n_{i = 1} \int_{0}^{\tau_i} A(t)^{\top} W_i(t)dN_i(t) \nonumber \\
    &= \frac{1}{n}\sum^n_{i = 1}\int_{0}^{\tau_i} A(t)^{\top} [\widetilde{W}_i(t) - \xi_i(t)][\widetilde{W}_i(t) - \xi_i(t)]^{\top} A(t)dt \nonumber \\
    &\textcolor{white}{=} \quad - \frac{2}{n}\sum^n_{i = 1} \int_{0}^{\tau_i} A(t)^{\top} [\widetilde{W}_i(t) - \xi_i(t)]dN_i(t)  \nonumber\\
    &=  \frac{1}{n}\sum^n_{i = 1}\int_{0}^{\tau_i} A(t)^{\top} \widetilde{W}_i(t) \widetilde{W}_i(t)^{\top} A(t)dt - \frac{2}{n}\sum^n_{i = 1} \int_{0}^{\tau_i} A(t)^{\top} \widetilde{W}_i(t)dN_i(t) \nonumber \\
    &\textcolor{white}{=} \quad - \frac{2}{n}\sum^n_{i = 1} \int_{0}^{\tau_i} A(t)^{\top} \xi_i(t) \widetilde{W}_i(t)^{\top} A(t)dt  + \frac{1}{n} \sum^n_{i = 1} \int_{0}^{\tau_i} A(t)^{\top} \xi_i(t) \xi_i(t)^{\top} A(t)dt  \nonumber \\
    &\textcolor{white}{=} \quad + \frac{2}{n}\sum^n_{i = 1} \int_{0}^{\tau_i} A(t)^{\top} \xi_i(t)dN_i(t) \nonumber \\
    &= \frac{1}{n}\sum^n_{i = 1}\int_{0}^{\tau_i} A(t)^{\top} \widetilde{W}_i(t) \widetilde{W}_i(t)^{\top} A(t)dt  - \frac{2}{n}\sum^n_{i = 1} \int_{0}^{\tau_i} A(t)^{\top} \widetilde{W}_i(t)dN_i(t) \nonumber\\
    &\textcolor{white}{=} \quad +  \frac{1}{n} \sum^n_{i = 1} \int_{0}^{\tau_i} \alpha_X(t)^{\top} \epsilon_i(t) \epsilon_i(t)^{\top} \alpha_X(t)dt \nonumber \\
    &= \widetilde{R}_n(A) +  \frac{1}{n} \sum^n_{i = 1} \int_{0}^{\tau_i} \alpha_X(t)^{\top} \epsilon_i(t) \epsilon_i(t)^{\top} \alpha_X(t)dt.
\end{align}

\section{Simulation algorithm}
\label{simulation_algorithme}

The following pseudo-algorithm present simulation process. For a patient $i$, initial values are defined as follow :
\begin{itemize}
    \item At the initiation :
    \begin{itemize}
        \item $D_i(0) = 0$ ;
        \item $X_i(0) \sim \mathcal{U}([min_X, max_X])$ with $min_X$ and $max_X$ a k-dimensional vector;
        \item $Z^1_i \sim  \mathcal{U}([min_{Z_1}, max_{Z_1}])$, $Z^2_i \sim  \mathcal{B}er(p_{Z_2})$ and $Z^2_i \sim  \mathcal{P}(\lambda_{Z_3})$
    \end{itemize}
    \item For $t = 0,...,10$
    \begin{itemize}
        \item $N_i(]t, t+1]) \sim \text{Non homogeneous Poisson process with intensity function }\delta D(t) + \delta_0 +\sum_{j=1}^{d_Z} \delta_{Zj} Z^j + \sum_{j=1}^{d}\delta_{Xj} X^j(t)$
        \begin{itemize}
        \item If $D_i(t) = 0$ :
        \begin{itemize}
            \item $X_i(t+1) = \kappa_{D0}X_i(t) + \epsilon$, with $\epsilon \sim \mathcal{N}(0_{\mathbb{R}^{d_X}},\Sigma)$
            \item $X^{0}_i(t+1) = X_i(t+1)$
            \item $D_i(t) \sim \mathcal{B}er(m \sum_{j=1}^{d_X}\lambda_j e^{\sum_{j=1}^{d_X} \lambda_j X_i^j(t)})$
        \end{itemize}
         \item If $D_i(t) = 1$ :
        \begin{itemize}
            \item $X_i(t+1) = X_i(t) + \kappa_{D1}(\sqrt{1000} - X_i(t) ) + \epsilon$, with $\epsilon \sim \mathcal{N}(0_{\mathbb{R}^{d_X}},\Sigma)$
            \item $X^0_i(t+1) = \kappa_{D0}X_i(t) + \epsilon$, with $\epsilon \sim \mathcal{N}(0_{\mathbb{R}^{d_X}},\Sigma)$
            \item $D_i(t+1) = 1$
 \end{itemize}
        \end{itemize}
    \end{itemize}
\end{itemize}

The following parameters were used for the simulations presented in the paper whatever the number of time-varying covariates studied : $min_{Z_1} = -20$,$max_{Z_1} = -10$, $p_{Z_2} = 0.5$, $\lambda_{Z_3} = 0.1$, $\delta = -0.01 $, $\delta_0 = 30$, $\delta_Z = (0.1,0.02,0.01)$, $m = 1.5$. Table~\ref{tab:parameter} presents parameter values depending of the number of simulated time-varying covariates.

\begin{table}[h]
    \centering
    \begin{tabular}{|l|c|c|c|}
    \hline
    Parameters & 1 covariate & 3 covariates  & 6 covariates \\
    \hline
      $min_{X}$ & $0$ & $(0, 0, 0)$ &  $(0, 0, 0, 0, 0, 0)$  \\
      \hline
      $max_{X}$ & $10$ & $(10, 20, 30)$ & $(10, 20, 30, 10, 20, 30)$  \\
      \hline
      $\kappa_{D0}$ & $-0.25$ & $(-0.25, -0.25, -0.25)$ & $(-0.25, -0.25, -0.25, -0.25, -0.25, -0.25)$  \\
      \hline
      $\kappa_{D1}$ & $0.25$ & $(0.25, 0.25, 0.25)$ &  $(0.25, 0.25, 0.25, 0.25, 0.25, 0.25)$ \\
      \hline
      $\delta_X$ & $-0.25$ & $(-0.3,0,-0.25)$ & $(-0.3, -0.2, 0, 0, -0.2, -0.25)$  \\
      \hline
      $\lambda $ & $0.12$ & $(0.16, 0.14, 0)$ & $(0.13, 0.12, 0.13, 0.14, 0, 0)$ \\
      \hline
    \end{tabular}
    \caption{Parameter values depending of the number of simulated time-varying covariates}
    \label{tab:parameter}
\end{table}

